\documentclass[10pt]{article}
\usepackage[letterpaper]{geometry}
\usepackage{hicss}
\usepackage{times}
\usepackage{amsfonts}
\usepackage{amsmath}
\usepackage{comment}
\usepackage[none]{hyphenat}
\usepackage{url}
\usepackage{latexsym}
\usepackage{minted}
\usepackage{indentfirst}
\usepackage{minted}
\usepackage[utf8]{inputenc}
\usepackage{textcomp}
\usepackage{newunicodechar}
\newunicodechar{−}{\textminus}
\usepackage{textcomp}
\usepackage{graphicx}
\usepackage{hyperref}
\usepackage{tikz}
\usepackage{multirow}
\usepackage{siunitx}
\usetikzlibrary{fit}

\usepackage{microtype}
\AtBeginDocument{\frenchspacing}

\usetikzlibrary{arrows.meta, positioning, shapes.geometric}

\usepackage{placeins}

\usetikzlibrary{positioning,calc}
\graphicspath{{images/}}
\usepackage[
    style=apa,
  ]{biblatex}
\newcommand{\generators}{\mathcal{G}}
\newcommand{\loads}{\mathcal{D}}
\newcommand{\lines}{\mathcal{L}}
\newcommand{\branches}{\mathcal{B}}

\title{Multi-Objective Deep Reinforcement Learning for Secure and Stable Power System Operation}

 \author{Ioannis Papadopoulos \\
  Department of Wind and Energy Systems \\
  Technical University of Denmark \\
  {\underline{ iopap@dtu.dk}}
  \And
  Georgios Tsaousoglou \\
  Department of Applied Mathematics\\and Computer Science \\
Technical University of Denmark \\
  {\underline{ geots@dtu.dk} }
  \AND
  Johanna Vorwerk \\
  Department of Wind and Energy Systems \\
  Technical University of Denmark \\
  {\underline{ vorjo@dtu.dk} }
  }

\date{}

\begin{document}
\maketitle
\begin{abstract}
The ongoing energy transition challenges the stable operation of power systems and increases the need for rapid decision-making under uncertainty. While reinforcement learning has emerged as a promising framework for power system control and operation, existing applications typically focus on a single operational criterion, such as thermal security or small-signal stability. However, power system operation is inherently multi-objective and may involve trade-offs between objectives.
This paper develops a unified-control deep reinforcement learning agent that maintains thermal security under stochastic load variations while steering the system toward operating points with improved damping of the most critical mode. Compared to a thermal-security-only agent and a business-as-usual policy, the proposed agent achieves a better balance among the operational objectives considered, with notably improved damping and negligible thermal-security violations. Finally, the operational value of increased critical damping is demonstrated under small- and large-signal disturbances, where operating points with higher damping lead to faster oscillation decay and improved critical clearing times.
\end{abstract}

\subsubsection*{Keywords:}

power system operation, deep reinforcement learning, thermal security, small-signal stability, transient stability

\section{Introduction}
The ongoing energy transition challenges power system operation. Increasing demand driven by sector-wide electrification, gradual decommissioning of dispatchable units, integration of intermittent renewable energy sources, and liberalization of power markets collectively push power systems closer to their operational limits and undermine their stable operation. 
In parallel, uncertainty in renewable generation and rapidly changing demand patterns render operating conditions more variable and less predictable, increasing the need for rapid decision-making and control.

The operational relevance of these challenges was recently illustrated by the 28 April 2025 blackout of the Iberian Peninsula. According to (\cite{ENTSOE_2025_Blackout}), the event was preceded by operator actions aimed at mitigating poorly damped oscillations. However, these actions appear to have interacted unfavorably with the operating point of the system, contributing to fast voltage increases, consequent cascading disconnection of generation assets across the entire Iberian Peninsula, and ultimately the blackout of continental Spain and Portugal.

This case provides evidence that established operational protocols may have detrimental impacts on the system if they remain agnostic to its current operating point. Moreover, it highlights that the complexity of modern power systems poses significant challenges for power system operators in accounting simultaneously for traditional security criteria, such as thermal security, and increasingly relevant stability aspects, such as small-signal stability.

In this context, (\cite{Capitanescu}) stresses the need for enhanced decision-support tools, while noting that computing optimal actions may require high-performance computing resources due to the multiple sources of computational complexity. More specifically, (\cite{TS_OPF}) observes that directly embedding dynamic constraints into conventional mathematical optimization formulations is computationally demanding and typically tractable only for small systems with simplified dynamic models. Extending such formulations to larger or more realistic networks generally requires further simplifying assumptions to ensure tractability, at the cost of reduced accuracy in representing system dynamics. Even when such formulations remain tractable, the resulting computational complexity is generally prohibitive for real-time deployment, in contrast to learning-based alternatives that concentrate the computational effort in an offline training phase and, once trained, produce decisions at very fast inference speeds (\cite{DiVito2024}). Jointly, these challenges motivate reinforcement learning (RL) as a suitable framework for decision-making and control in power systems, with prior work identifying it as a viable approach for a broad range of power system control problems (\cite{RL_review}).

\subsection{Related Work}
RL-based control for the secure operation of power systems from a static perspective has been explored in the context of the Learning to Run a Power Network competitions. In the 2020 challenge, agents were trained to maintain a secure electricity supply while alleviating thermal overloadings caused by adversarial actions or stochastic variations in generation and demand profiles (\cite{L2RPN2020}). Similar concepts were adopted in the 2023 version, where the action space was extended to include renewable-curtailment and storage-utilization actions (\cite{L2RPN2023}). However, these approaches primarily attempt to maintain feasible operation with respect to thermal overloadings and do not assess any stability aspects. Therefore, the selected operating points may be thermally secure without necessarily satisfying other stability criteria.

RL-based approaches have also been proposed for stability enhancement. At the device level, (\cite{MultiBandPSS}) develops an agent for tuning the parameters of a multi-band power system stabilizer, showing improved damping of the targeted modes under different operating conditions and short-circuit faults. At the system-wide level, (\cite{WideAreaDamping}) coordinates multiple local damping controllers to achieve a prescribed minimum damping for critical modes. Similarly, (\cite{PhysicsInformedDRL}) computes a time-varying feedback gain matrix for selected synchronous generators to improve modal damping under large-signal disturbances, while incorporating graph-based representations to support operation under topological variations. However, these damping-oriented approaches do not explicitly enforce operational constraints, such as N-0 and N-1 thermal security. Moreover, although the referenced damping-control studies consider different initial operating points, the load levels are generally fixed during each episode, so the agents are not trained under persistent stochastic load variations.

Despite these advances, sequential decision-making agents that simultaneously account for static security and dynamic stability have not been studied, to the best of the authors' knowledge. Decision-making in power systems has always been multi-objective, requiring trade-offs across static-security and dynamic-stability requirements. This complexity has traditionally been managed by studying static and dynamic phenomena separately, with dedicated tools for each class. However, the energy transition is challenging this paradigm, as the growing number of controllable resources and the uncertainty introduced by renewable generators require decisions to be taken within drastically reduced time frames, and jeopardize the assumption that static and dynamic phenomena can be studied independently. 
Therefore, addressing these challenges requires the design of agents capable of selecting actions that remain acceptable across coupled static-security and dynamic-stability requirements under time-varying operating conditions.

\subsection{Contributions}
This paper contributes toward the development of decision-making agents for power systems that consider multiple static security and dynamic stability criteria within a unified decision-making framework. Specifically, it investigates whether a deep reinforcement learning (DRL) agent can maintain N-0 and N-1 thermal security under stochastic load variations while also steering the system toward operating points with improved damping of the most critical system mode. The proposed framework enables both preventive and corrective actions, as the agent can act before violations occur and respond once they emerge. Particular emphasis is placed on small-signal stability because it relates directly to the new stability classes, such as converter-driven and resonance stability, introduced by converter-based resources (\cite{New_Stability}). To this end, a unified-control DRL agent is developed and compared against two reference policies, namely a thermal-security-only agent and a business-as-usual (BAU) policy, representing the case in which no actions are taken in the absence of N-0 or N-1 thermal security violations. The comparison is performed under unseen stochastic load trajectories to assess both thermal-security performance and damping improvement. Finally, an additional case study highlights how improved critical damping achieved by the unified-control agent translates into enhanced system stability under small-signal and large-signal disturbances. 

\section{Learning Framework for Unified Security and Stability Control}

\subsection{Unified-Control Agent Architecture}
\begin{figure}[t]
    \centering
    \begin{tikzpicture}[
        every node/.style = {font=\small},
        block/.style = {
            rectangle,
            rounded corners=4pt,
            draw=black,
            thick,
            minimum width=2.5cm,
            minimum height=1.0cm,
            align=center,
            fill=gray!10,
        },
        agent/.style = {
            circle,
            draw=black,
            thick,
            minimum size=1.9cm,
            align=center,
            fill=gray!15,
        },
        env/.style = {
            block,
            fill=gray!15,
        },
        arrow/.style = {
            -{Stealth[length=2.5mm, width=2.5mm]},
            thick,
        },
        reward arrow/.style = {
            -{Stealth[length=2.5mm, width=2.5mm]},
            thick,
            dashed,
        },
        side label/.style = {
            font=\small,
            align=center,
            inner sep=1pt,
        },
    ]
        \node[agent] (agent) {Unified-control\\agent};
        \node[env, below=1.0cm of agent] (env) {Power system};

        \coordinate (leftBottom)  at ([xshift=-1.0cm]env.west);
        \coordinate (leftTop)     at (leftBottom |- agent.west);
        \coordinate (rightTop)    at ([xshift=1.0cm]agent.east);
        \coordinate (rightBottom) at (rightTop |- env.east);

        \draw[arrow]
            (env.west) -- (leftBottom)
            -- coordinate[midway] (stateMid) (leftTop)
            -- coordinate[pos=0.25] (rewardTap) (agent.west);

        \node[side label, left=2pt of stateMid, anchor=east, font=\footnotesize]
        {\textbf{State}\\Topology\\Dispatch\\Demand\\Voltages\\Loadings
        };

        \coordinate (rewardCorner) at ([yshift=0.84cm]rewardTap);
        \draw[reward arrow]
            (rewardTap) -- (rewardCorner)
            -- coordinate[midway] (rewardMid)
            (agent.north west);

        \node[font=\small, above=2pt of rewardMid, anchor=south, xshift=-5pt] {Reward signal
        };
        
        \draw[arrow]
            (agent.east) -- (rightTop)
            -- coordinate[midway] (actionMid) (rightBottom)
            -- (env.east);

        \node[side label, right=2pt of actionMid, anchor=west, font=\footnotesize]          {\textbf{Action}\\Line Switching\\Redispatch\\Voltage Setpoints\\Idle
        };

    \end{tikzpicture}
    \caption{Interaction between the unified-control agent and the power system.}
    \label{fig:rl_loop}
\end{figure}
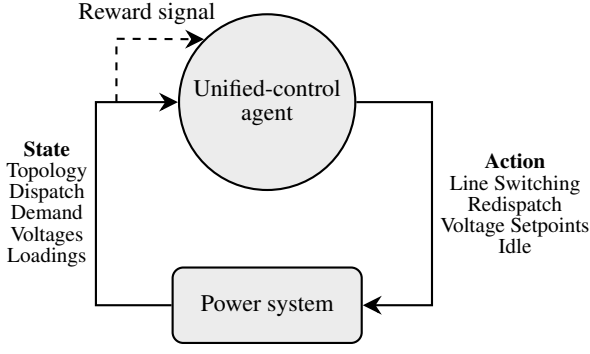

In this subsection, a visual representation of the proposed agent is provided. During deployment, the DRL-based unified-control agent operates according to the closed-loop process shown in Figure \ref{fig:rl_loop}. At each control step, the power system communicates its current operating state to the agent, whose embedded neural network outputs an action intended to maintain thermally secure operation while improving damping. The action is applied to the system, where it interacts with any external disturbance acting simultaneously (e.g., load variation), producing a new operating point that is fed back to the agent.

\subsection{Introduction to Markov Decision Processes and Deep Reinforcement Learning}
Decision-making under uncertainty is commonly modeled using a Markov Decision Process (MDP). The tuple~$\mathcal{M}$ defines an MDP:
\begin{equation}
    \mathcal{M} = \left(\mathcal{S},\, \mathcal{A},\, p, \,r, \,\gamma \right),
\end{equation}
where $\mathcal{S}$ denotes the state space, $\mathcal{A}$ denotes the action space, $p(s_{t+1}\mid s_t, a_t)$ governs the transition dynamics from state $s_t$ to state $s_{t+1}$ after applying action $a_t$, while $r_t = r(s_t,a_t,s_{t+1})$ denotes the reward received by the agent and $\gamma$ is a discount factor. 

RL provides a framework for solving MDPs through interaction with the environment. At each step, the agent observes the state, selects an action according to a policy $\pi(a \mid s)$, transitions to a new state, and receives a reward. 
The objective is to learn a policy that maximizes the expected discounted return. DRL extends RL by using neural networks as function approximators to learn an approximate solution to the MDP. 

A widely used DRL algorithm is the Proximal Policy Optimization (PPO) (\cite{PPO}). PPO belongs to the broader class of policy gradient methods and is known for its stable training behavior, mainly achieved by constraining the size of each policy update through a clipping mechanism. In PPO, the actor neural network represents the policy $\pi_\theta(a \mid s)$, while the critic neural network approximates the state-value function $V^\pi(s)$ through $V_\phi(s)$, where $\theta$ and $\phi$ are trainable parameters. The state-value function can be written using the Bellman equation:

\begin{equation}
V^\pi(s_t) = \mathbb{E}\left[r_t + \gamma V^\pi(s_{t+1}) \mid s_t\right].
\end{equation}

During training, the critic is fitted so that $V_\phi(s)$ approximates the discounted-return targets, while the actor is updated to increase the probability of actions that lead to higher observed returns. Through these iterative updates, PPO produces an approximate solution to the MDP in the form of a learned policy.

\subsection{Problem Formulation for Unified Security and Stability Control}

In this subsection, the unified-control problem is formulated as an MDP by defining its state space, action space, and reward function.

\paragraph{State and Action Spaces}
We consider a power system comprising a set $\generators$ of generators, a set $\loads$ of loads, a set $\lines$ of transmission lines, and a set $\branches$ of branches, including transmission lines and transformers. The index $g=0$ denotes the slack generator, while $\generators_{-0}$ denotes the set of non-slack generators. 

The state vector $s_t$ (i.e., the agent's observation) consists of the binary transmission line status vector $b_{i,t}$, with $i \in \lines$, the generator active power setpoints $P_{g,t}$, with $g \in \generators$, the active and reactive load demand values $P_{d,t}$ and $Q_{d,t}$, with $d \in \loads$, the voltage setpoints $V_{g,t}$ of the non-slack generators, with $g \in \generators_{-0}$, and the branch loading levels $\ell_{k,t}$, with $k \in \branches$.
\begin{equation}
    s_t =
    \left[
    b_{i,t},\,
    P_{g,t},\,
    P_{d,t},\,
    Q_{d,t},\,
    V_{g,t},\,
    \ell_{k,t}
    \right].
    \label{eq:state_vector}
\end{equation}

The branch loading levels $\ell_{k,t}$ are expressed as percentages of their nominal thermal limits. Since the agent is not allowed to modify transformer statuses, these are not included in the branch status vector $b_{i,t}$.

The action vector $a_t$ is discrete and includes transmission line switching actions $u_{i,t}$, active power redispatch actions $\Delta P_{g,t}$ for the non-slack generators, voltage setpoint adjustment actions $\Delta V_{g,t}$ for the non-slack generators, and an idle action:
\begin{equation}
\begin{aligned}
    a_t \in
    \left\{
    u_{i,t},\,
    \Delta P_{g,t},\,
    \Delta V_{g,t},\,
    \mathrm{idle}
    \right\}, \\
    i \in \mathcal{L},\quad g \in \generators_{-0}.
\end{aligned}
\label{eq:action_space}
\end{equation}

\paragraph{Thermal Security Criterion}
The thermal-security criterion assesses whether branch loading levels remain within their nominal thermal limits. In the base case, corresponding to N-0 operation, the loading of each branch is evaluated at the current operating point. For the N-1 assessment, single transmission line contingencies are considered by removing one transmission line at a time and evaluating the resulting post-contingency branch loadings. Branches whose loading exceeds their nominal thermal limit are counted as thermal overloadings. This criterion therefore captures both existing overloadings and overloadings that may arise after a single transmission line outage.

\paragraph{Small-Signal Stability Criterion}
The small-signal stability criterion selected in this work is based on the modal properties of the system around the current operating point. More specifically, the nonlinear dynamic model is linearized around its operating point, yielding a state-space representation of the form:

\begin{equation}
    \Delta \dot{x}=A\Delta x,
\end{equation}
where $x$ contains the dynamic state variables and $A$ is the state matrix. The eigenvalues $\lambda=\sigma \pm j\omega$ of $A$ characterize the local dynamic modes of the system. The real part $\sigma$ indicates whether a mode decays or grows over time, while the imaginary part $\omega$ determines its oscillation frequency. For an oscillatory mode, the damping ratio is computed as:

\begin{equation}
    \zeta = \frac{-\sigma}{\sqrt{\sigma^2+\omega^2}}.
    \label{eq:damping_ratio}
\end{equation}

A higher damping ratio indicates faster decay of oscillations after a disturbance, whereas negative damping corresponds to an unstable oscillatory mode. The damping ratio is therefore used as an indicator of small-signal stability, since it directly quantifies how effectively oscillations are damped. The relevant quantity is the damping ratio of the least-damped mode, as this critical mode decays most slowly and therefore governs the system's small-signal stability margin. Accordingly, improving overall system damping amounts to acting on this critical mode and shifting it further into the left half-plane.

\paragraph{Reward Function}
The reward function is designed to guide the unified-control agent toward thermally secure operating points while accounting for operational costs and small-signal stability. In more detail, in \eqref{eq:reward_damping_aware}, the terms $n_{0,t}$ and $n_{1,t}$ denote the number of N-0 and N-1 thermal overloadings, respectively, while $\zeta_t$ denotes the damping ratio of the system's least-damped mode. These quantities are computed at each state according to the thermal-security and small-signal-stability criteria described above. The term $r_{\mathrm{surv}}$ denotes the survival reward assigned when the power flow (PF) calculation converges. If the PF calculation does not converge, the episode is terminated and the terminal penalty $R_{min}$ is assigned. Episodes are not terminated upon the detection of negative damping, so that the agent can learn to recover from such undesired operating conditions.

The operational cost terms are defined as follows. The redispatch cost $c_P(a_t)$ is proportional to the absolute magnitude of active power redispatch, while the voltage control cost $c_V(a_t)$ is proportional to the absolute voltage setpoint adjustment. The slack generator penalty $\psi(P_{0,t})$ is proportional to the violation magnitude of the slack generator active power limits. These cost terms are given by:

\begin{subequations}
\begin{align}
    c_P(a_t)
    &=
    \sum_{g \in \generators_{-0}}
    \alpha_P \left|\Delta P_{g,t}\right|,
    \label{eq:redispatch_cost}\\
    c_V(a_t)
    &=
    \sum_{g \in \generators_{-0}}
    \alpha_V \left|\Delta V_{g,t}\right|,
    \label{eq:voltage_cost} \\
\psi(P_{0,t})
&=\alpha_0
\big[
\max(0, \underline{P}_0 - P_{0,t}) \\ \nonumber
&\quad \quad +\max(0, P_{0,t} - \overline{P}_0)
\big],
\label{eq:slack_penalty}
\end{align}
\end{subequations}
where $\alpha_P$, $\alpha_V$, and $\alpha_0$ are the redispatch cost, voltage control cost, and slack penalty coefficients, respectively. The parameters $\underline{P}_0$ and $\overline{P}_0$ denote the lower and upper active power limits of the slack generator.

The mathematical formulation of the reward function for the unified-control agent is given in \eqref{eq:reward_damping_aware}. The coefficients $\lambda_0$, $\lambda_1$, and $\lambda_{\zeta}$ are weighting factors associated with N-0 thermal overloadings, N-1 thermal overloadings, and damping, respectively.

\begin{equation}
\begin{aligned}
R_t =
\begin{cases}
&-\lambda_0 n_{0,t}
-\lambda_1 n_{1,t}
+\lambda_{\zeta}\zeta_t
+r_{\mathrm{surv}} \\
&\quad -c_P(a_t)- c_V(a_t) \\
&\quad -\psi(P_{0,t}),
\quad \text{if PF converges},\\
&R_{\min},
\quad \text{otherwise}.
\end{cases}
\end{aligned}
\label{eq:reward_damping_aware}
\end{equation}

\section{Experimental Implementation}

\subsection{Case Studies}
In this work, the proposed unified-control agent, which accounts for both thermal security and small-signal stability, is compared against two reference policies, namely a thermal-security-only agent and the BAU policy. The thermal-security-only agent is obtained from the unified-control agent by setting the damping-related reward coefficient $\lambda_{\zeta}$ in (\ref{eq:reward_damping_aware}) equal to zero. All other agent characteristics, including the state space, action space, hyperparameters, and training principles, remain unchanged. Finally, the BAU policy remains idle in the absence of N-0 or N-1 thermal security violations. 

\subsection{Test System and Dynamic Modeling}
The proposed framework is applied to the 9-bus system, depicted in Figure~\ref{fig:wscc9_sld}. This system is selected as a proof-of-concept platform, since the main objective of this work is to investigate whether a multi-objective unified-control agent can be effectively trained and to quantify the operational gains achievable relative to the reference policies. Scaling to larger systems and reducing training cost are deferred to future work. Static data are sourced from (\cite{WSCC9}), while the formulation of all dynamic components is taken from (\cite{PowerSystems}). Synchronous generators are represented using the simplified Marconato four-state synchronous-machine model. Each generator is further equipped with an automatic voltage regulator of type AVRTypeI and a turbine governor of type SteamTurbineGov1. Importantly, loads are modeled as constant admittances.

\begin{figure}[t]
\centering
\resizebox{\columnwidth}{!}{%
\begin{tikzpicture}[every node/.style={font=\small}]
    \tikzstyle{busbar}=[line width=1.6pt]
    \tikzstyle{branch}=[line width=0.8pt]
    \tikzstyle{gen}=[draw, circle, minimum size=8mm]
    \tikzstyle{loadarrow}=[->, line width=0.8pt]

    \coordinate (b1) at (0,0);
    \coordinate (b4) at (0,2.0);

    \coordinate (b5) at (-2.0,3.4);
    \coordinate (b6) at (2.0,3.4);

    \coordinate (b7) at (-2.0,5.0);
    \coordinate (b8) at (0,5.0);
    \coordinate (b9) at (2.0,5.0);

    \coordinate (b2) at (-4.0,5.0);
    \coordinate (b3) at (4.0,5.0);

    \coordinate (b7_low) at (-2.0,4.8);
    \coordinate (b7_elbow) at (-1.5,4.8);
    \coordinate (b5_right) at (-1.5,3.4);
    \coordinate (b4_left) at (-1.5,2);

    \coordinate (b9_low) at (2.0,4.8);
    \coordinate (b9_elbow) at (1.5,4.8);
    \coordinate (b6_left) at (1.5,3.4);
    \coordinate (b4_right) at (1.5,2);

    \draw[busbar] ($(b1)+(-0.40,0)$) -- ($(b1)+(0.40,0)$);
    \draw[busbar] ($(b4)+(-2.0,0)$) -- ($(b4)+(2.0,0)$);

    \draw[busbar] ($(b5)+(-0.80,0)$) -- ($(b5)+(0.80,0)$);
    \draw[busbar] ($(b6)+(-0.80,0)$) -- ($(b6)+(0.80,0)$);

    \draw[busbar] ($(b7)+(0,-0.45)$) -- ($(b7)+(0,0.45)$);
    \draw[busbar] ($(b8)+(0,-0.45)$) -- ($(b8)+(0,0.45)$);
    \draw[busbar] ($(b9)+(0,-0.45)$) -- ($(b9)+(0,0.45)$);

    \draw[busbar] ($(b2)+(0,-0.45)$) -- ($(b2)+(0,0.45)$);
    \draw[busbar] ($(b3)+(0,-0.45)$) -- ($(b3)+(0,0.45)$);

    \draw[branch] (b7) -- (b8);
    \draw[branch] (b8) -- (b9);

    \draw[branch] (b7_low) -- (b7_elbow);
    \draw[branch] (b7_elbow) -- (b5_right);
    \draw[branch] (b5_right) -- (b4_left);

    \draw[branch] (b9_low) -- (b9_elbow);
    \draw[branch] (b9_elbow) -- (b6_left);
    \draw[branch] (b6_left) -- (b4_right);

    \draw[branch] (b1) -- (0,0.72);
    \draw[branch] (0,0.93) circle (0.22);
    \draw[branch] (0,1.18) circle (0.22);
    \draw[branch] (0,1.40) -- (b4);

    \draw[branch] (b2) -- (-3.22,5.0);
    \draw[branch] (-3.00,5.0) circle (0.22);
    \draw[branch] (-2.75,5.0) circle (0.22);
    \draw[branch] (-2.53,5.0) -- (b7);

    \draw[branch] (b9) -- (2.53,5.0);
    \draw[branch] (2.75,5.0) circle (0.22);
    \draw[branch] (3.00,5.0) circle (0.22);
    \draw[branch] (3.22,5.0) -- (b3);

    \node[gen] (G1) at (0,-0.85) {$G_1$};
    \node[gen] (G2) at (-4.85,5.0) {$G_2$};
    \node[gen] (G3) at (4.85,5.0) {$G_3$};

    \draw[branch] (G1.north) -- ($(b1)+(0,-0.02)$);
    \draw[branch] (G2.east) -- ($(b2)+(-0.02,0)$);
    \draw[branch] ($(b3)+(0.02,0)$) -- (G3.west);

    \draw[loadarrow] ($(b5)+(0,-0.03)$) -- ($(b5)+(0,-0.72)$);
    \draw[loadarrow] ($(b6)+(0,-0.03)$) -- ($(b6)+(0,-0.72)$);
    \draw[loadarrow] ($(b8)+(0.03,0.22)$) -- ($(b8)+(0.72,0.22)$);

    \node[align=center] at (-2.0,2.55) {Load 1};
    \node[align=center] at (2.0,2.55) {Load 2};
    \node[align=left] at (0.95,5.55) {Load 3};

    \node[above] at ($(b2)+(0,0.52)$) {2};
    \node[above] at ($(b3)+(0,0.52)$) {3};

    \node[right] at ($(b1)+(0.52,0)$) {1};
    \node[right] at ($(b4)+(2.08,0)$) {4};

    \node[left]  at ($(b5)+(-0.9,0)$) {5};
    \node[right] at ($(b6)+(0.9,0)$) {6};

    \node[above] at ($(b7)+(0,0.52)$) {7};
    \node[above] at ($(b8)+(0,0.52)$) {8};
    \node[above] at ($(b9)+(0,0.52)$) {9};

    \node at (0.48,1.055) {$T_1$};
    \node at (-2.88,5.45) {$T_2$};
    \node at (2.88,5.45) {$T_3$};

\end{tikzpicture}%
}
\caption{Single-line diagram of the 9-bus test system.}
\label{fig:wscc9_sld}
\end{figure}

\subsection{Training Episodes}
\label{subsec:TrainingScenarios}
Training of the agents is performed under episodes with stochastic load variations, which constitute the external disturbances of the MDP. An episode is defined as a sequence initialized in a common reset state $s_0$ and consists of 20 state transitions, each driven by a set of external load deviations and an agent action. These variations are intended to represent both gradual changes and sudden net load deviations, for instance due to abrupt changes in renewable generation or demand. More specifically, for each of the system's loads, the episode generator produces an independent stochastic trajectory of 20 load deviations $\delta_t$, where $\delta_t$ denotes the relative deviation of the load from its nominal admittance value reflected in the state $s_t$. Each trajectory is initialized as $\delta_0 = 0$ at the reset state.

At each subsequent step, $\delta_t$ evolves according to the random-walk model in \eqref{eq:random_walk}, where $\rho_t$ is the persistence coefficient, $\epsilon_t$ is a Gaussian noise term, and $K_t$ is a possible large-jump term:

\begin{equation}
\delta_t = \delta_{t-1} + \rho_t(\delta_{t-1} - \delta_{t-2}) + \epsilon_t + K_t.
\label{eq:random_walk}
\end{equation}

The noise term follows $\epsilon_t \sim \mathcal{N}(0, 0.02^2)$. The persistence coefficient is set to $\rho_t = 0.25$, except immediately after a large jump, when $\rho_t = 0$ to avoid propagating the jump-induced change through the momentum term. A large jump can occur only for $t = 5,\ldots,12$, with probability $0.2$ at each eligible state, provided that no previous jump has occurred in the same trajectory. This placement exposes the agent to operating conditions before and after the sudden large load deviation, allowing it to learn both preventive and corrective actions. The large-jump term $K_t$ is defined by \eqref{eq:jump_magnitude}, where $q_t$ is a random sign sampled uniformly from $\{-1,+1\}$, and $X_t$ is the absolute jump magnitude sampled from $\mathcal{N}(0.5,0.1^2)$ and truncated using $\max(0,X_t)$:

\begin{equation}
K_t =
\begin{cases}
q_t \max(0, X_t), & \text{if a jump is triggered,} \\
0, & \text{otherwise.}
\end{cases}
\label{eq:jump_magnitude}
\end{equation}

Thus, the large jump is centered around $\pm 0.5$. At most one large jump is allowed per load trajectory, and the load trajectories are sampled independently using the same parameter values. The numerical values of the random-walk parameters are selected empirically, since representative real load-deviation patterns were not available. Overall, 60~000 episodes are generated, of which 500 are set aside for evaluation.

\begin{figure}[t]
    \centering
    \includegraphics[width=\columnwidth]{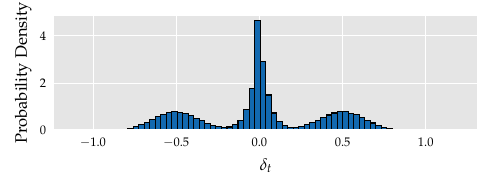}
    \caption{Distribution of the generated load deviation values $\delta_t$ for all three loads across the generated episodes.}
    \label{fig:delta_distribution}
\end{figure}

Figure \ref{fig:delta_distribution} illustrates the distribution of the generated $\delta_t$ values for all three loads of the 9-bus system across the 60~000 episodes. A significant probability mass is concentrated around zero, reflecting the small stochastic load variations introduced by the random-walk component described in \eqref{eq:random_walk}. Additionally, probability mass appears symmetrically around $\pm 0.5$, which is consistent with the large-jump mechanism defined in \eqref{eq:jump_magnitude}. Overall, the distribution confirms that the episode generator produces both small load fluctuations and occasional larger deviations in either direction.

\subsection{Action Space and Reward Function Parametrization}
Regarding the action space, for $g \in \generators_{-0}$, the active power redispatch actions are discretized as:
\begin{equation}
    \Delta P_{g,t}
    \in
    \{-50,\,-45,\,-40,\,\ldots,\,40,\,45,\,50\}\text{ MW },
    \label{eq:redispatch_set}
\end{equation}
subject to the active power limits of generator $g$. The selected redispatch granularity and range allow the agent to perform fine-tuning actions, but also rapidly modify the generator dispatch within a few states. Similarly, the voltage control actions are discretized as:
\begin{equation}
    \Delta V_{g,t}
    \in
    \{-0.05, -0.04,\ldots, 0.04, 0.05\}\text{ p.u.},
    \label{eq:voltage_set}
\end{equation}
with the resulting voltage setpoint constrained to avoid system-wide under-voltage and over-voltage conditions:
\begin{equation}
    0.95 \leq V_{g,t} \leq 1.05.
    \label{eq:voltage_limits}
\end{equation}

Table~\ref{tab:reward_coefficients} reports the numerical values of all coefficients of the reward function. These values are determined through manual reward shaping to produce agents that respect thermal-security constraints (i.e., N-0 and N-1), while leaving sufficient gradient signal on the remaining reward components (i.e., damping, action cost, slack regulation) to differentiate policies.

\begin{table}[t]
    \centering
    \scriptsize
    \sffamily
    \renewcommand*{\arraystretch}{1.3}
    \caption{Reward function coefficients.}
    \label{tab:reward_coefficients}
    \begin{tabular}{lcc}
        \hline
        Coefficient & Unit & Value \\
        \hline
        $\lambda_0$ & -- & 25 \\
        $\lambda_1$ & -- & 0.5 \\
        $\lambda_{\zeta}$ & -- & 20 \\
        $r_{\mathrm{surv}}$ & -- & 1 \\
        $\alpha_P$ & cost/MW & 0.02 \\
        $\alpha_V$ & cost/p.u. & 20 \\
        $\alpha_0$ & cost/MW & 0.5 \\
        $R_{\min}$ & -- & -10~000 \\
        \hline
    \end{tabular}
\end{table}

The training and evaluation environment is implemented in Julia using the PowerSystems.jl (\cite{PowerSystems}) and PowerSimulationsDynamics.jl (\cite{PSD}) packages. The Julia-based PF simulator defines the transition dynamics of the MDP by computing the post-action operating point. The DRL agents are trained in Python, and the policy is updated using the thermal-security, damping, and operational-cost metrics returned by the Julia environment.

\subsection{Deployed Agent Model}
The formulated MDP is solved using the PPO algorithm, leveraging the Maskable PPO model from the Stable Baselines3 Contrib library (\cite{Maskable_PPO}). A key feature of this model is that, during training, the agent observes the state vector, and action masking prevents the selection of physically infeasible actions. For example, in this application, the masking mechanism prevents the agent from opening lines that are already open, redispatching generators beyond their nominal limits, or violating the prescribed voltage control ranges. In this way, the use of Maskable PPO simplifies the reward function and avoids the inclusion of additional penalty terms for infeasible actions.

The model's adopted hyperparameters are summarized in Table~\ref{tab:ppo_hyperparameters}. These include the discount factor $\gamma$, the learning rate $\eta$, and the entropy coefficient $\epsilon$. The chosen configuration proves effective in handling the complexity of the posed problem, as evidenced by the training curves and final performance assessment presented in the following sections. Additionally, for both the unified-control and thermal-security-only approaches, the agents are trained using five different seeds to assess the consistency of the obtained results. Consequently, 10 DRL agents are exported at the end of the training process. During training, a single action is allowed per transition, and each agent is trained for a total of 500~000 action steps.

\begin{table}[t]
    \centering
    \scriptsize
    \sffamily
    \renewcommand*{\arraystretch}{1.3}
    \caption{Adopted hyperparameters.}
    \label{tab:ppo_hyperparameters}
    \begin{tabular}{lcc}
        \hline
        Hyperparameter & Symbol & Value \\
        \hline
        Discount factor & $\gamma$ & 0.99 \\
        Learning rate & $\eta$ & 3e-4 \\
        Entropy coefficient & $\epsilon$ & 0.02 \\
        Number of training steps & -- & 500~000 \\
        Random seeds & -- & a, b, c, d, e \\
        \hline
    \end{tabular}
\end{table}

\subsection{Training Curves}
\begin{figure}[t]
    \centering
    \includegraphics{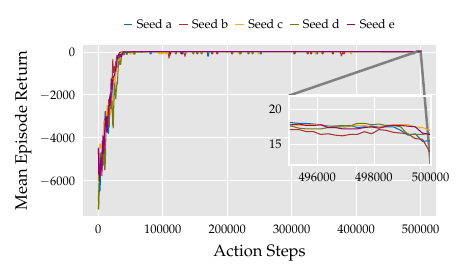}
    \caption{Mean episode return during training for thermal-security-only agents across five seeds.}
    \label{fig:reward_DU}
\end{figure}

\begin{figure}[t]
    \centering
    \includegraphics{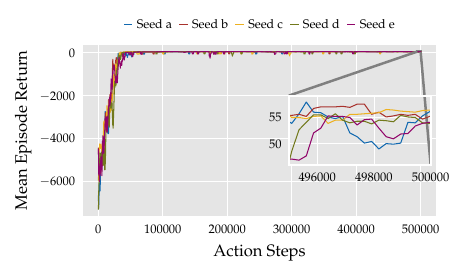}
    \caption{Mean episode return during training for unified-control agents across five seeds.}
    \label{fig:reward_DA}
\end{figure}

Figures \ref{fig:reward_DU} and \ref{fig:reward_DA} show the training curves for the thermal-security-only and unified-control agent configurations, respectively. 
It can be observed that in the early training phase, policies trigger terminal conditions, thus producing episode returns in the order of −5~000 to −7~000 due to the large terminal penalty $R_{min}$. However, as all 10 agents subsequently explore the action space, they learn to avoid these terminal conditions, as indicated by the sharp increase in the mean episode return. By the end of the training process, all agents converge to policies that consistently yield positive episode returns, as indicated by the plateau of the curves. The inset zooms in on approximately the final 5~000 action steps. Across all seeds, for a training process incorporating 500~000 action steps, the average training time was 15.06 hours. 

\section{Results}

\subsection{Policies Evaluation Principles}
All three policies, namely the unified-control agent, the thermal-security-only agent, and the BAU policy, are evaluated on the same set of 500 unseen load trajectories, as described in Subsection \ref{subsec:TrainingScenarios}. For a fair comparison of returns across policies, the damping reward coefficient $\lambda_{\zeta}$ is set to 20 in the evaluation environment, even though the thermal-security-only agents do not receive a damping-related reward during training.

\subsection{Policies Comparison}
Table \ref{tab:mean_episodic_return} summarizes the mean episode return over the 500 evaluation episodes, per policy and per seed. The reported return should not be interpreted as a scalar RL score alone, but as an aggregate measure of operational performance derived from metrics with direct physical meaning. In this sense, higher returns indicate a more effective balance among the considered operational objectives. In more detail, the BAU policy yields a negative mean return, while both trained agent configurations achieve strongly positive returns, demonstrating that learned control actions provide a better solution to the multi-objective combinatorial problem considered in this work. Among the trained agents, the unified-control policies achieve the highest returns across all seeds, indicating that the additional small-signal-stability objective is effectively integrated into the control problem. Importantly, across all 50~000 step-level evaluations of the unified-control agents, no PF failures or operating points with negative damping are observed.

\begin{table}[t]
    \centering
    \scriptsize
    \sffamily
    \renewcommand*{\arraystretch}{1.3}
    \caption{Mean episode return across the 500 evaluation episodes.}
    \label{tab:mean_episodic_return}
    \begin{tabular}{lcc}
        \hline
        Policy & Seed & Mean Episode Return \\
        \hline
        \multirow{5}{*}{Thermal-security-only}
        & a  & 51.36 \\
        & b  & 52.35 \\
        & c  & 53.03 \\
        & d & 52.48 \\
        & e & 51.21 \\
        \hline
        \multirow{5}{*}{Unified-control}
        & a  & 58.85 \\
        & b  & 58.81 \\
        & c  & 57.62 \\
        & d & 56.79 \\
        & e & 56.56 \\
        \hline
        Business-as-usual & -- & -7.22 \\
        \hline
    \end{tabular}
\end{table}

\begin{figure}[t]
    \centering
    \includegraphics{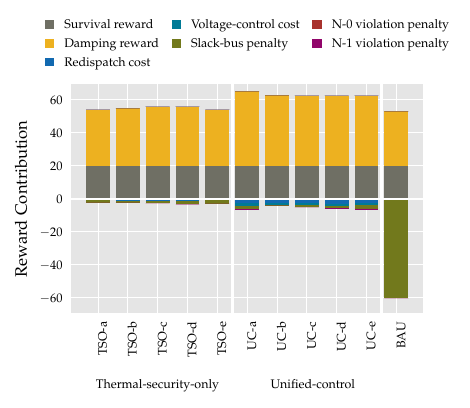}
    \caption{Mean per-episode reward-component decomposition across the 500 evaluation episodes for each thermal-security-only and unified-control seed and for the business-as-usual policy.}
    \label{fig:fig_reward_breakdown}
\end{figure}

The decomposition of the reward's components in Figure \ref{fig:fig_reward_breakdown} further clarifies the origin of the performance differences among the policies. The BAU policy's negative mean return is driven primarily by violations of the slack generator's active power limits, since the policy does not redispatch the non-slack units to compensate for stochastic load variations. In contrast, the thermal-security-only agents largely eliminate these violations through generator redispatch, incurring a moderate redispatch cost in return. The positive damping reward observed under both the BAU and the thermal-security-only policies is a passive side effect of the system's intrinsic dynamics and is not actively pursued by either policy. 

\begin{figure}[t]
    \centering
    \includegraphics{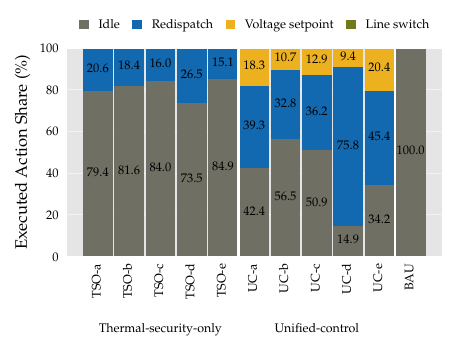}
    \caption{Executed action shares across the 500 evaluation episodes for each thermal-security-only and unified-control seed and for the business-as-usual policy.}
    \label{fig:fig_actions}
\end{figure}

The unified-control agents also limit slack violations effectively, but they additionally accumulate substantially higher damping reward, which accounts for their consistently superior performance across all seeds. 
Importantly, both trained agent configurations produce negligible N-0 and N-1 thermal violations throughout the 500 evaluation episodes, with the maximum identified mean penalties across all seeds being 0.05 and 0.021 for N-0 and N-1, respectively. 

The action-frequency distributions in Figure~\ref{fig:fig_actions} provide additional insights. Both thermal-security-only and unified-control agents rely on generator redispatch to enforce thermal security and slack-limit compliance, but only the unified-control agents exploit voltage setpoint control. This behavior is consistent with the system operator's intuition that voltage control, rather than redispatch, is the preferred lever for improving the system's damping. Therefore, this observation suggests that the unified-control agents learn a control strategy that aligns with established operator practice.

\begin{figure}[t]
    \centering
    \includegraphics{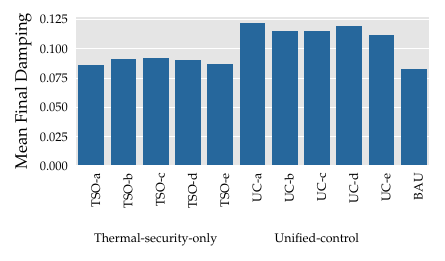}
    \caption{Mean final damping ratio across the 500 evaluation episodes.}
    \label{fig:mean_final_damping}
\end{figure}

Finally, Figure \ref{fig:mean_final_damping} reports the mean end-of-episode damping ratio across the 500 evaluation episodes. The unified-control agents achieve markedly higher final damping than both the BAU policy and the thermal-security-only agents. As previously mentioned, the latter two converge to damping levels that depend primarily on the stochastic load realizations rather than on any active control choice.

\subsection{Operational Impact of Improved Damping}

To highlight the operational benefits of higher damping, the transient response of the system is assessed under load-step disturbances and symmetrical faults. The simulations are initialized from operating points with different damping levels, including high-damping points obtained from the unified-control agent and medium- to low-damping points obtained from the BAU policy.
The BAU policy is selected as the baseline for this comparison because, as shown in Figure~\ref{fig:mean_final_damping}, it yields operating points with lower average damping than the thermal-security-only agents, thereby providing a meaningful low-damping reference.

\begin{figure}[t]
    \centering
    \includegraphics{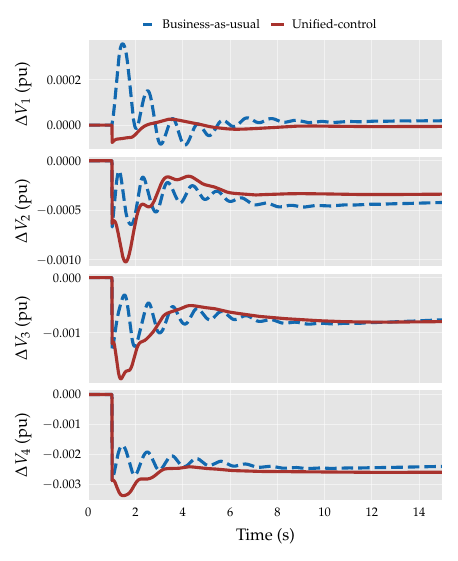}
    \caption{Terminal voltage deviations following a load step at bus 8.}
    \label{fig:terminal_voltages}
\end{figure}

Across the 500 evaluation episodes, we focus on the episode where the unified-control agent (i.e., UC-a) achieves the highest damping, corresponding to 15.37\%. For the same scenario, the BAU policy converges to an end-episode damping ratio of 7.78\%. From these two operating points, a load step is applied at load bus 8. Assuming constant voltage magnitudes, this load step corresponds to a 50\% increase in the load consumption at bus 8 at that operating point. Figure \ref{fig:terminal_voltages} depicts the voltage deviations $\Delta V$ at the three generator terminals and at load bus 8. It can be observed that, when the system is operated at increased damping, the oscillations induced by the load step are damped faster, thereby reducing the risk of instability.

\begin{figure*}[!t]
    \centering
    \includegraphics[width=\textwidth]{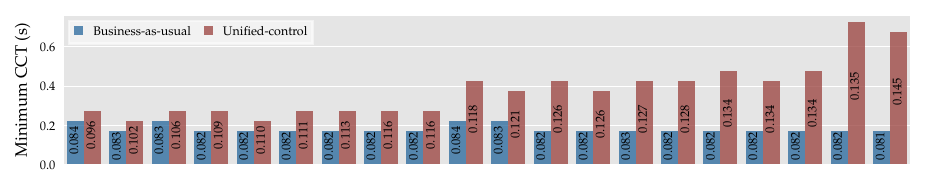}
    \caption{Minimum critical clearing times obtained at operating points reached by the business-as-usual policy and the unified-control agent, with episodes sorted by increasing final damping under the unified-control agent.}
    \label{fig:ranked_minimum_CCT_Mid_comparison}
\end{figure*}

Extending the analysis to large-signal stability, the final operating points of 20 episodes, obtained under the BAU policy and the unified-control agent, are evaluated in terms of the system's minimum critical clearing time (CCT). More specifically, three-phase faults, modeled as fixed admittances, are applied to all high-voltage buses, and a CCT scan is performed. Each operating point is assigned the minimum CCT identified across all considered fault locations. Instability is detected if at least one rotor-angle separation with respect to the slack machine exceeds $360^\circ$ at any time instant during the dynamic simulation.

Figure~\ref{fig:ranked_minimum_CCT_Mid_comparison} compares the minimum CCTs obtained at the operating points reached by the unified-control agent and the BAU policy, with episodes sorted by increasing final damping under the unified-control agent. Among all operating points, those obtained by the unified-control agent exhibit higher damping, with the minimum damping in this set being 9.63\%, compared to a maximum damping of 8.41\% across the operating points obtained by the BAU policy. It can be seen that across all episodes, the operating point with the improved damping is more robust to large-signal disturbances, as evidenced by the consistently higher CCT values. Moreover, although the relationship is not strictly proportional, the results show a clear tendency for higher damping to coincide with improved CCT values.

\section{Conclusion}
This work investigated the development of a deep reinforcement learning agent capable of balancing multiple operational objectives, including thermal security and small-signal stability, under stochastic load variations. The results show that the proposed unified-control agent outperforms the reference policies, indicating that the adopted reward design and training procedure enable the agent to coordinate thermal-security requirements with small-signal stability improvement. This represents a step beyond existing approaches, which typically address these security and stability aspects separately, and supports the development of control agents that account for multiple operational criteria within a unified decision-making framework. The operational value of steering the system toward high-damping operating points was further demonstrated under both small- and large-signal disturbances. In particular, the unified-control agent achieved markedly higher damping levels, which translated into faster oscillation decay and increased critical clearing times. Overall, the presented case studies on the 9-bus system clearly demonstrate that multi-objective agents that jointly consider static and dynamic operational performance enhance system robustness by producing operating points that are simultaneously more secure and more stable.

Future work will focus on improving the action-space formulation by integrating continuous actions and allowing simultaneous combinations of actions within each decision step. In addition, the applicability and scalability of the proposed approach to larger systems will be examined.







\FloatBarrier
\printbibliography

\end{document}